# Light-Induced Magnetic Precession in (Ga,Mn)As Slabs: Hybrid Standing-Wave Damon-Eshbach Modes

D. M. Wang,[1] Y. H. Ren,[1] X. Liu,[2] J. K. Furdyna,[2] M. Grimsditch,[3] and R. Merlin[1]

[1]*FOCUS Center and Department of Physics, The University of Michigan, Ann Arbor, MI 48109-1040*

[2]*Department of Physics, University of Notre Dame, Notre Dame, IN 46556-5670*

[3]*Materials Science Division, Argonne National Laboratory, Argonne, IL 60439*

*September 25, 2006*

Coherent oscillations associated with spin precessions were observed in ultrafast optical experiments on ferromagnetic (Ga,Mn)As films. Using a complete theoretical description of the processes by which light couples to the magnetization, values for the exchange and anisotropy constants were obtained from the field-dependence of the magnon frequencies and the oscillation-amplitude ratios. Results reveal a relatively large negative contribution to the energy due to surface anisotropy leading to excitations that are a mixture of bulk waves and surface modes.

Following the pioneering work of Ohno et al. on (Ga,Mn)As [1] and (In,Mn)As [2], and the subsequent discovery of high-temperature ferromagnetism in (Ga,Mn)N [3], ferromagnets based on III-V compounds have been the subject of intense study prompted, in part, by the promise that their compatibility with semiconductor technology may lead to novel nonvolatile spintronic applications [4]. Although it is now well established that the ferromagnetism in these alloys arises from the exchange interaction between holes and the manganese ions [5] and, hence, that the Curie temperature ($T_C$) is determined mainly by the hole concentration [6,7,8], the relationship between transport, magnetic properties and defect configurations is not yet fully understood and, moreover, important magnetic parameters remain poorly known. In this letter, we present an investigation of the spin dynamics of (Ga,Mn)As using subpicosecond laser pulses to generate and probe coherent magnetization precessions [9,10,11,12,13]. We also discuss a comprehensive model of the magnetic eigenmodes and their coupling to light, which allows one to gain accurate values of the exchange, bulk and surface anisotropy constants from time-domain data, establishing a hitherto unrecognized relationship between nonlinear optics and ferromagnetic resonance (FMR).

We studied several (Ga,Mn)As films, grown by low-temperature molecular beam epitaxy on semi-insulating (001) GaAs substrates. Their parameters are listed in Table I. Using a Ti-sapphire laser that provided ~ 70 fs pulses of central wavelength 800 nm at the repetition rate of 82 MHz, pump-probe experiments were performed in the reflection geometry at ~ 4-6 K. The laser beams, of energy density per pulse ~ 0.2 µJ/cm$^2$ (pump) and 0.03 µJ/cm$^2$ (probe), penetrated the slabs along the z–axis [001]. The pump pulses induce coherent magnetic precessions modifying the reflection of the probe pulses that follow behind. We measured the pump-induced change in the field of the reflected probe beam, $\delta\mathbf{E}_R$, also known as the coherent scattered field, as a function of the time delay between the two pulses. To get $\delta\mathbf{E}_R$, we determined, both, the pump-induced shift of the polarization angle of the reflected probe field, $\delta\theta$, and the differential reflectivity $\delta R \propto |\mathbf{E}_R+\delta\mathbf{E}_R|^2 - |\mathbf{E}_R|^2 \approx 2\mathbf{E}_R \cdot \delta\mathbf{E}_R$ ($\mathbf{E}_R$ is the reflected probe field when the pump is turned off). $\delta\theta$ was obtained from differential magnetic Kerr (DMK) measurements in the Voigt geometry using a scheme that gives an output signal proportional to $\mathbf{E}_R \times \delta\mathbf{E}_R$ [14]. Because scattering by spin-flip excitations is described by an antisymmetric tensor [15], the signature of a pure-spin magnetic precession is $\delta\mathbf{E}_R \perp \mathbf{E}_R$ (or $\delta R \equiv 0$). This selection rule was found to be strictly obeyed in all cases.

To calculate the magnetic modes, we consider a (Ga,Mn)As film occupying the region $|z| \leq L/2$. The Landau-Lifshitz equation of motion for the magnetization $\mathbf{M}$ is [16]

$$\partial \mathbf{M}/\partial t = \gamma \mathbf{M} \times \left[ \mathbf{H} - \nabla_\mathbf{M} F_A + DM^{-1}\nabla^2 \mathbf{M} \right] \quad (1)$$

where $F_A = -K_2 M_z^2/M^2 - K_4\left(M_x^4 + M_y^4\right)/2M^4 - K_O(M_x - M_y)^2/2M^2$ is the magnetic anisotropy contribution to the free-energy density [17,18]. $K_2$, $K_4$ and $K_O$ are constants and $\mathbf{H}$ is the field strength. The term $\propto K_O$, accounting for a surface reconstruction of orthorhombic symmetry present in the films [19], leads to a monotonic temperature dependence of the orientation of the equilibrium-magnetization $\mathbf{M}_0$ from near [100] at $T = 0$ to [110] for $T \geq 30$K (at $\mathbf{H} = 0$) [19,20]. While this behavior is ultimately the reason why light pulses can trigger coherent pre-cessions (see later), the $K_O$-term introduces a very small correction to the frequency and intensity of the eigenmodes and, thereby, it will be momentarily ignored. Assuming that both **H** and **M** depend only on $z$, we get using the Maxwell equations of magnetostatics $H_y = 0$ and $H_z + 4\pi M_z = 0$ [21]. In our experiments, the external field and, hence, **M**$_0$ are parallel to the easy-axis which, for simplicity, we assume to be along [100] [19]. Decomposing the field into static and microwave components, $\mathbf{H} = H_0\mathbf{i} + \mathbf{h}(\mathbf{r},t)$ with $|\mathbf{h}| \ll H_0$, we get from (1) $\mathbf{M} \approx (M_0, m_y \sin\Omega t, m_z \cos\Omega t)$ where $|m_{y,z}| \ll M_0$ and

$$(\Omega/\gamma)m_y = [H_0 + 4\pi M_0 + 2M_0^{-1}(K_4 - K_2) - D\partial^2/\partial z^2]m_z$$
$$(\Omega/\gamma)m_z = (H_0 + 2M_0^{-1}K_4 - D\partial^2/\partial z^2)m_y \qquad (2)$$

Thus, the solutions to (1) are either bulk-like standing waves ($\sin\alpha z$ and $\cos\alpha z$) or surface states ($\sinh\alpha z$ and $\cosh\alpha z$) similar to those introduced by Damon and Eshbach (DE) for $F_A = D = 0$ [22]. The eigenfrequencies are

$$\Omega^2 = \Omega_C^2 \pm D\alpha^2\left[\gamma^2\left(H_0 + 2K_4/M_0\right) + \Omega_C^2/\left(H_0 + 2K_4/M_0\right)\right] + \gamma^2 D^2\alpha^4 \qquad (3)$$

where $\Omega_C^2 = \gamma^2[H_0 + 2K_4/M_0][H_0 + 4\pi M_0 + 2M_0^{-1}(K_4 - K_2)]$ and the plus (minus) sign corresponds to the bulk- (surface-) like modes. From (3), it can be shown that the four solutions are all of the DE-type for $\Omega < \Omega_C$ whereas two of them are surface- and the other two bulk-like for $\Omega > \Omega_C$. It follows that the precessions for $\Omega > \Omega_C$ are generally of mixed bulk-surface character. The actual eigenmodes are selected by the conditions at the boundaries which, in turn, depend on the surface energy density $F_S$. Here, we take $F_S = K_S M_z^2/M_0^2$ (same for the two surfaces; therefore, the solutions are either even or odd with respect to $z = 0$) and use the Rado-Weertman condition $\mathbf{M} \times (\nabla_\mathbf{M} F_S - DM_0^{-1} \partial \mathbf{M}/\partial n) = 0$ [23] giving $\partial m_y/\partial z = 0$ and $\partial \ln m_z/\partial z = \mp 2K_S/DM_0$ at $z = \pm L/2$; $\mathbf{n}$ is a unit vector normal to the layers. We note that these conditions cannot be met by a standing wave. Hence, they are incompatible with the well-known Kittel's pinning condition, $m_y = m_z = 0$ [24], which only applies if $\mathbf{M}_0$ is not in-plane [25]. It is important to emphasize the crucial role played by the surface in determining the slab's magnetization pattern and the precession eigenfrequencies. In Fig. 1, we show calculations for the two lowest even modes, $S_0$ and $S_2$ (which, as shown below, are the only ones that are relevant to our experiments) of a film of width $L = 71$ nm. These results illustrate the progression from almost-pinned to unpinned standing waves, as we move from large positive values of $K_S$ to zero anisotropy, as well as the DE (bulk-surface) character of $S_0$ ($S_2$) for $K_S < 0$. The hybrid nature of the modes brings into question previous analyses of the available FMR data, based on Kittel's [26] or the unpinned ($K_S = 0$) condition [27], which led to the conclusion that the magnetization is not distributed uniformly along the direction of growth.

The experimental results are summarized in Figs. 2 and 3. The top graph in Fig. 2 shows DMK data at $T = 4$K for the as-grown 120-nm-thick sample after subtraction of an exponentially decaying background which reflects the spin relaxation of photoexcited electrons [28]. Consistent with the fact that they affect δθ but not δR, the oscillations are assigned to the precession of the magnetization around $\mathbf{M}_0$. We used linear prediction methods to fit the time-domain data and, as shown in the bottom panel of Fig. 2, the Fourier transform of the fit reveals two modes, which are assigned to $S_0$ and $S_2$. With the exception of the thinnest film ($L = 25$ nm), these two modes were observed in the DMK spectra of all the samples. The dependence of their frequencies on the applied magnetic field is shown in Fig. 3, together with FMR results for the $L = 71$ nm slab. The bottom graph gives the dependence of the ratio between the amplitudes of the two oscillations, $A_2/A_0$. As we shall see, this ratio is critical to the determination of the magnetic parameters, particularly the surface constant $K_S$.

The coupling between magnetic precessions and probe pulses is controlled by the spin-flip Raman susceptibility, which is closely related to that for Faraday rotation [15]. In our Voigt geometry, the scattered probe field associated with the $m_y$-component of the precession is polarized along the $z$-axis and, therefore, gives no DMK signal. Using results for scattering by coherent vibrations [29,30], we get for $m_z$-scattering

$$\left(\nabla^2 - \frac{n_R^2}{c^2} \times \frac{\partial^2}{\partial t^2}\right)(e_x \pm ie_y) = \frac{4\pi\chi_M}{c^2} \times \frac{\partial^2}{\partial t^2}[m_z(z,t)(e_y \pm ie_x)] \qquad (4)$$

where $\mathbf{e} = (e_x, e_y, 0)$ is the probe electric field, $n_R$ is the refractive index, $\chi_M = \partial\chi^{(0)}/\partial M$ and $\chi^{(0)}$ is the linear susceptibility. Let us define the average $\langle m_z(t) \rangle = (1/L)\int_{-L/2}^{+L/2} m_z(z,t)dz$. To lowest order, and provided (i) $Ln_R/c \ll 2\pi/\Omega$, (ii) the light penetration depth is $\gg L$ and (iii) multiple reflections can be ignored (these provisos are well obeyed in our experiments), it can be shown that coherent scattering is equivalent to a slowly-varying modulation of the refractive index

$$\delta n_R(t) = \pm(2\pi\chi_M/n_R)\langle m_z(t) \rangle \qquad (5)$$

with different signs for the two senses of circular polarization. Except for the constant factors, this expression is identical to that describing FMR. Because $\langle m_z(t) \rangle$ vanishes for odd preces-sions, (5) supports our contention that the two modes observed in the experimental data are $S_0$ and $S_2$.

In order to extract the magnetic parameters from the experimental data, we still need to ascertain the mechanism by which photoexcitation leads to precessions. Unlike the sharp selection rules of Raman type observed in probe scattering, we find that the strength of the oscillations is nearly the same for pump pulses of arbitrary circular or linear polarization. Since the Raman coupling for the pump operates only if the beams are circularly polarized [12,13], such a mechanism cannot be the cause of the precessions in our case [31]. Instead, our results point towards a relatively simpler thermal origin which, as first identified in ferromagnetic metals [10,11], relies on the temperature dependence of the anisotropy. Following the model proposed by van Kampen et al. [10], we attribute the magnetic coherence to a (in the scale of the magnon period) sudden deviation of the orientation of $\mathbf{M}_0$ due to the temperature rise that follows the absorption of the light pulse [32]. Since the equilibrium orientation, initially along [100], moves towards [110] after the pulse hits, and the light penetration depth is $\gg L$ so that the excitation is uniform across the film, this gives an effective coupling proportional to $\langle m_y \rangle$. Combining this result with (5), the amplitude of the DMK signal for a given precession is then $\delta\theta \propto \langle m_y \rangle \cdot \langle m_z \rangle / \langle m_y^2 \rangle$. In Fig. 4, we show the surface-anisotropy dependence of the frequencies of the two lowest even modes and $A_2/A_0$. Because the amplitude ratio depends strongly on $K_S$ but is far less sensitive to the other magnetic constants, the observation of $S_2$ is, in itself, a strong indication that $K_S < 0$. The parameters in Table I were obtained from fits to the experimental data using the theoretical expression for δθ and Eq. (3). Our values for the bulk anisotropy are in fairly good agreement with FMR results [17, 26,27].

In summary, we observed coherent precessions of combined bulk and surface character and determined the exchange and anisotropy constants of (Ga,Mn)As using a comprehensive model of the coupling of the magnetization to light.

This work was supported by the NSF Focus Physics Frontier Center and by NSF-DMR 0603752.

prediction [30] that the Raman mechanism (also referred to as the inverse Faraday effect in the case of magnons) is the dominant source of coherence in transparent media.

[32] An estimate based on the available data [19] gives ~ 150 µK for the change required at 4 K to shift this angle by 1 µradian. These values are consistent with the parameters of our experiments and the thermal properties of low-temperature-grown GaAs.

## Table Caption

TABLE I. Manganese concentration ($x$), thickness ($L$) and Curie temperature ($T_C$) of the (Ga,Mn)As samples used in this work and values of the magnetic exchange ($D$), surface ($K_S$) and bulk ($K_2$ and $K_4$) anisotropy constants inferred from the pump-probe data. Unless noted, the films are as-grown specimens.

| Sample | $x$ | $L$ (nm) | $T_C$ (K) | $D$ (T.nm$^2$) | $2K_S/M_0$ (T.nm) | $2K_4/M_0$ (T) | $4\pi M_0 - 2K_2/M_0$ (T) |
|---|---|---|---|---|---|---|---|
| 1[a] | 0.03 | 120 | 80 | 25±4 | -2.4±0.6 | 0.07±0.02 | 0.44±0.05 |
| 2 | 0.03 | 120 | 70 | 18±3 | -2.0±0.5 | 0.07±0.02 | 0.42±0.05 |
| 3 | 0.05-0.06 | 71 | 65 | 5.8±1 | -1.3±0.4 | 0.08±0.02 | 0.50±0.05 |
| 4 | 0.05-0.06 | 47 | 65 | 5.8±1 | -1.1±0.3 | 0.09±0.02 | 0.50±0.05 |
| 5 | 0.05-0.06 | 25 | 65 | | | 0.09±0.04 | 0.50±0.05 |

[a] Sample annealed at 280° C for 1 h.

Figure Captions

FIGURE 1 (color online). Precession eigenstates for the two lowest even modes of a 71-nm-thick (Ga,Mn)As film, centered at $z = 0$, using the values of the parameters $D$, $K_2$ and $K_4$ listed in Table I and $2K_S/DM_0 = -0.33$, 0 and 1 nm$^{-1}$. The in-plane external field $H_0 = 0.1$ T is along [100].

FIGURE 2 (color online). Results at $T = 4$K for the as-grown sample with $L = 120$ nm. The top panel shows Voigt-geometry DMK data (open circles) at $H_0 = 0.17$ T. The applied field is parallel to [100]. The black curve is the linear prediction fit, which gives two modes of period 99.7 ps and 83.9 ps. Their contributions to the fitted signal are shown, respectively, by the blue and red curves. The bottom panel shows the Fourier transform of the fit and, in the inset, the calculated $m_z$-component for the three lowest eigenmodes, using the parameter values listed in Table I.

FIGURE 3 (color online). Measured magnetic-field dependence of the precession mode frequencies (top panel) and the ratio $A_2/A_0$ (bottom panel). For clarity, results for the films with $L = 120$ nm are not shown. Solid curves are fits using the theoretical expressions discussed in the text. The FMR spectrum of the 71-nm-thick sample is shown in the inset.

FIGURE 4 (color online). Calculated surface-anisotropy dependence of the frequency and the ratio between the amplitudes of the two lowest even modes for $L = 71$ nm at $H_0 = 0$. Values for other parameters are listed in Table I.

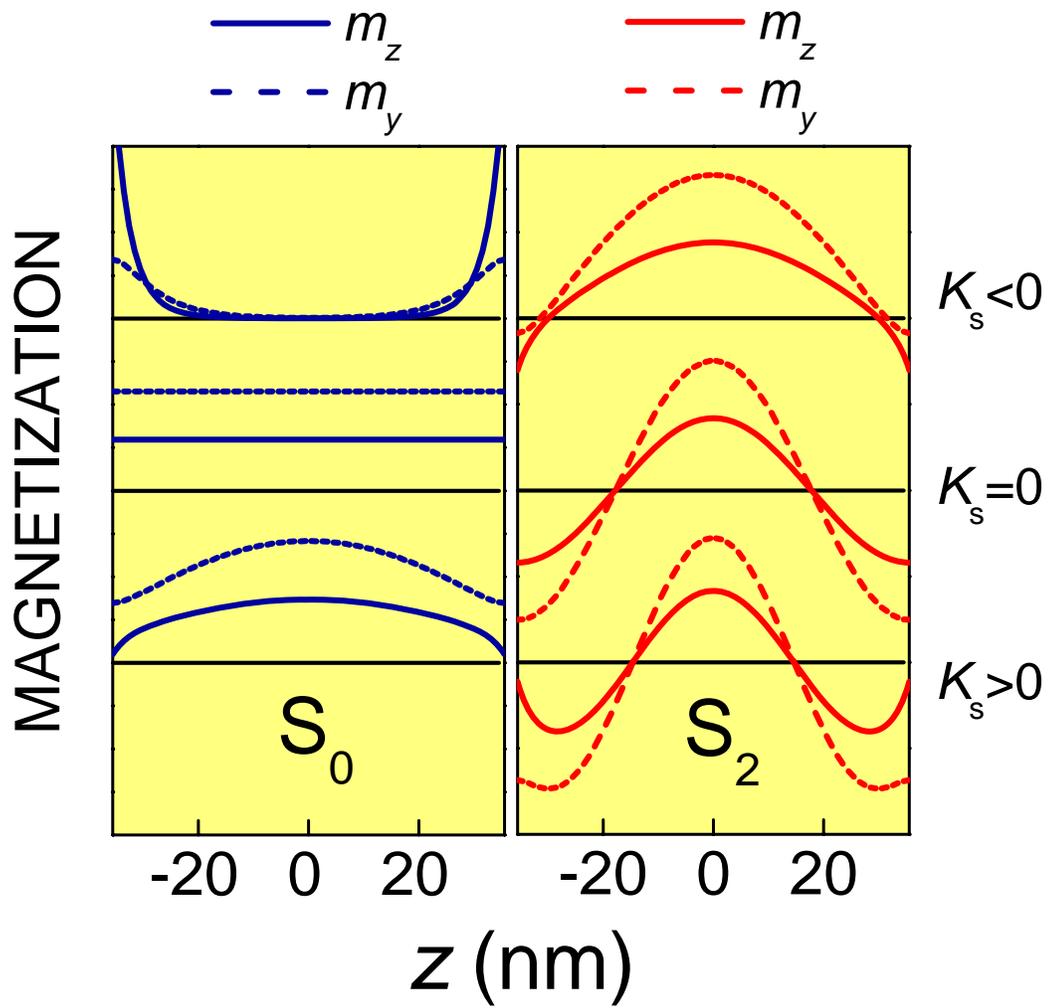

Figure 1

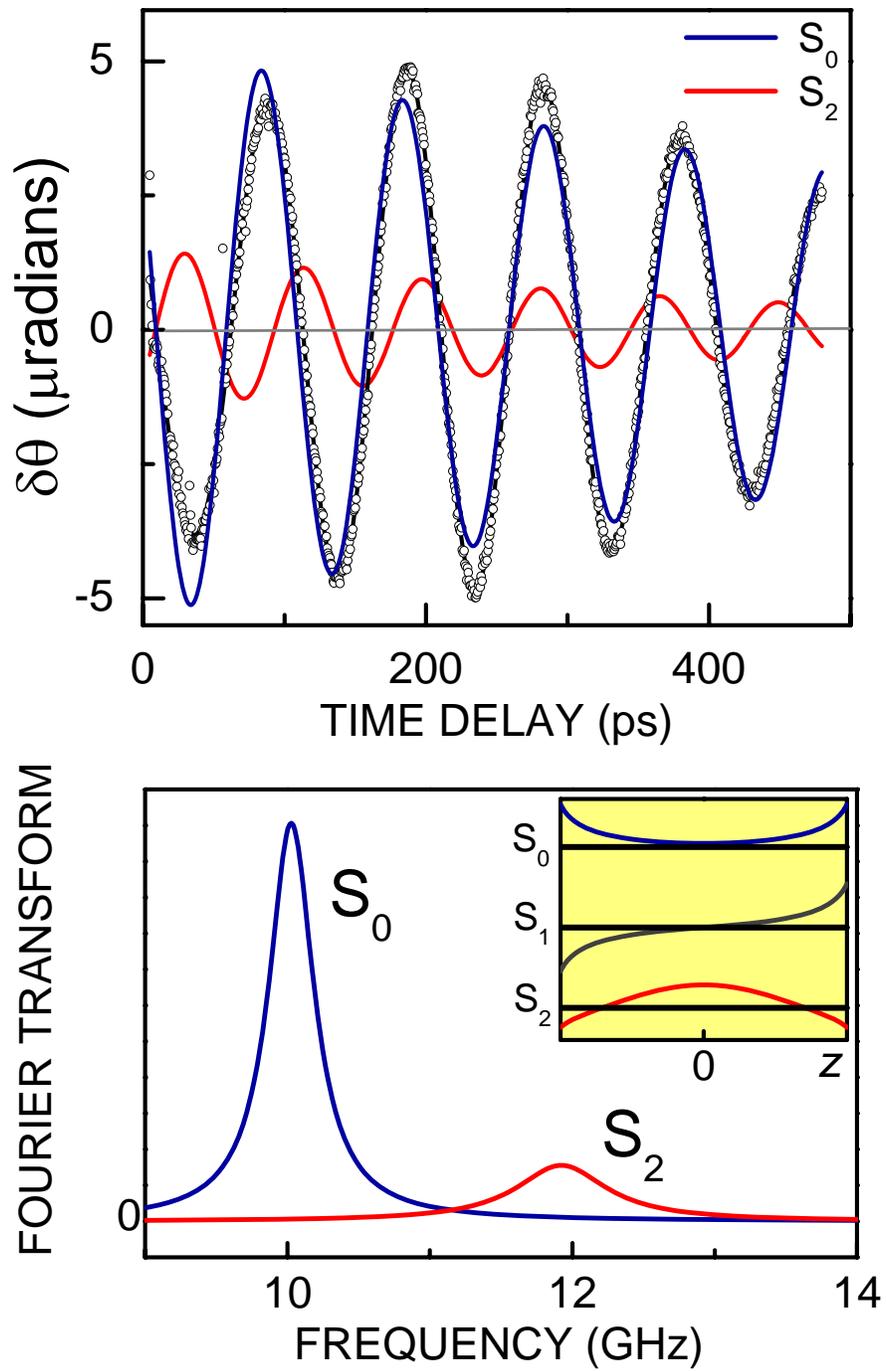

Figure 2

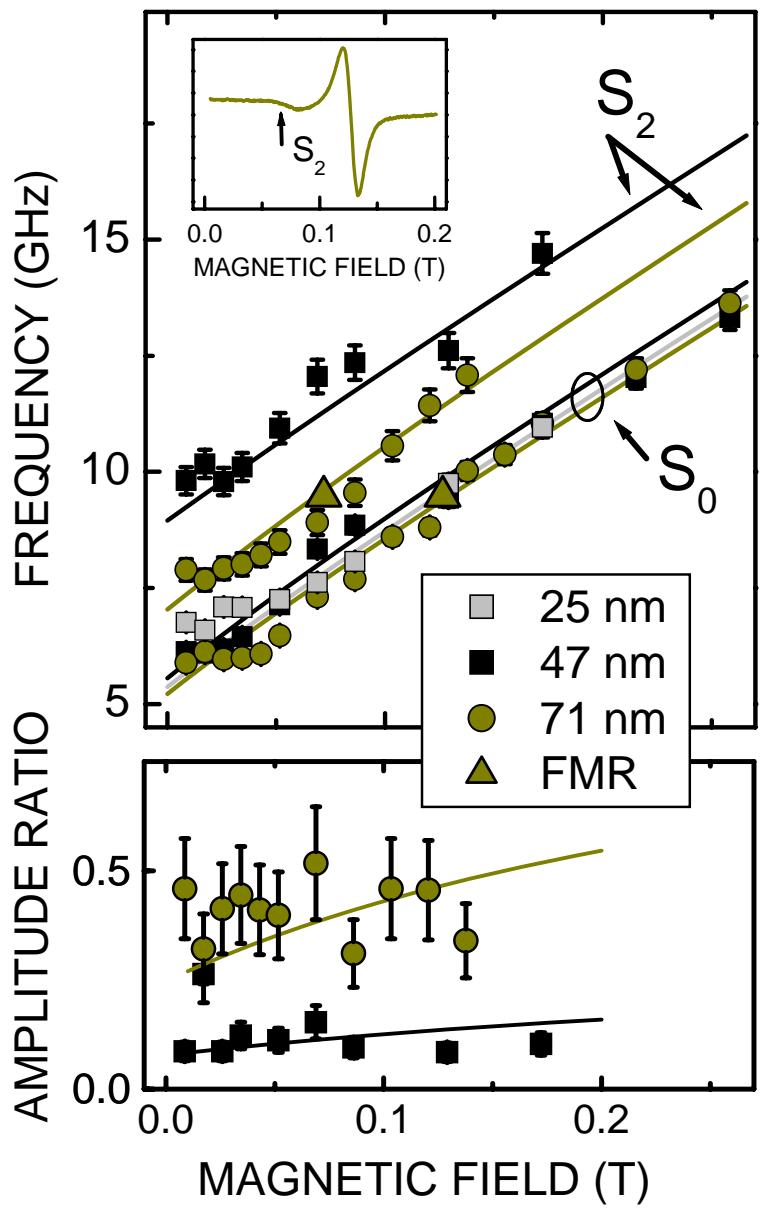

Figure 3

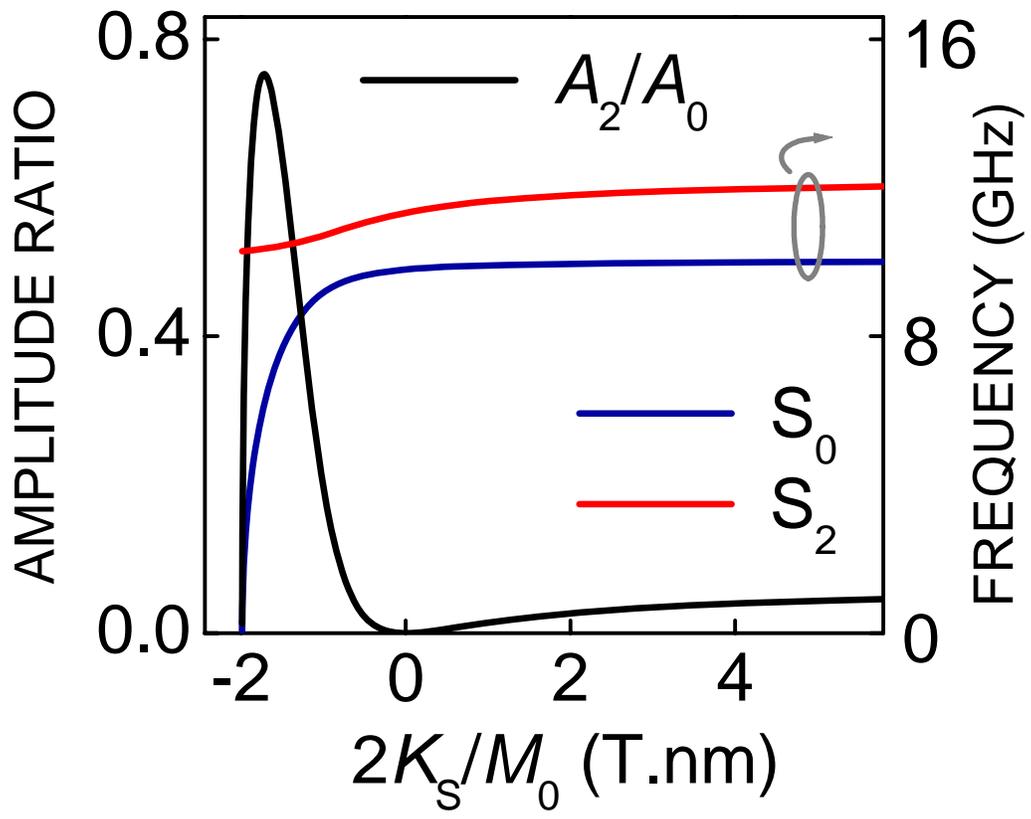

Figure 4